# CAPABILITIES OF NUMERICAL SIMULATION OF MULTIPHASE FLOWS IN CENTRIFUGAL PUMPS USING MODERN CFD SOFTWARE

**ALEXEY N. KOCHEVSKY**

*Research Scientist, Department of Applied Fluid Mechanics,
Sumy State University,
Rimsky-Korsakov str., 2, 40007, Sumy, Ukraine
alkochevsky@mail.ru*

**Abstract:** The paper describes capabilities of numerical simulation of liquid flows with solid and/or gas admixtures in centrifugal pumps using modern commercial CFD software packages, with the purpose to predict performance curves of the pumps treating such media. In particular, the approaches and multiphase flow models available in the package CFX-5 are described; their advantages and disadvantages are analyzed.

## Introduction

The problem of pumping of multiphase fluids, i.e., liquids with solid and/or gas admixtures, is urgent for many branches of industry. For pumping of multiphase fluids, even when low capacities and high heads are required, centrifugal pumps are usually used. The matter is, displacement pumps require high purity of the fluids pumped and exhibit low reliability when admixtures are present. Among the centrifugal pumps, radial-flow, mixed-flow or axial-flow pumps (in dependence on the required specific speed) of common design [1] have the highest efficiency. However, a disadvantage of these pumps is significant drop of parameters when the fluid pumped contains some gas and rapid clogging of the flow passage when some solid particles are present. As it is shown in [2, 3], for treating of multiphase fluids with high solid and/or gas contents, in dependence on the specific speed, free-vortex pumps and radial-flow pumps with low number of blades can be successfully used.

When the fluid pumped contains high volume of gas, the performance parameters of a pump fall down due to rupture of the fluid continuum by gas cavities [4]. When high volume of solid particles is present, the performance parameters collapse due to clogging of the flow passage of the pump [5]. The maximum content of admixtures that still does not cause interruption of pumping process depends on the composition of admixtures, design of the pump and its operation mode. In case of breakdown, the hydraulic network will also become out of order, and sometimes it may lead to damages far exceeding the cost of the pump equipment. For preventing failure of multiphase hydraulic networks, as a rule, the pumps are equipped with special additional units. However, use of such units causes increase of cost and decrease of reliability of these networks.

Taking this into account, in order to ensure reliable operation of the pumping equipment, a designer should be able to predict performance curves of the pumps and their change in dependence on the composition of the fluid pumped, before manufacturing of the pump. However, the theory of designing of centrifugal pumps was developed mostly for pumping of clean water. When pumping a multiphase fluid, the performance curves of pumps of different design vary in very different ways, and it is very difficult to predict performance parameters of a pump at different composition of the fluid without laborious experimental research.

During recent decades, methods of computational fluid dynamics (CFD) were developed and implemented in modern software packages and widely used for numerical simulation of fluid flows. The paper [6] presents contemporary approach for analysis of fluid flows in hydromachines. In the paper [7], capabilities of modern CFD software packages for modeling of complex physical processes in fluid flows are described, including also multiphase flows. Nowadays, the leading software packages, in particular, CFX, Fluent and STAR-CD, are able to simulate many complex flows, with high agreement with experimental data.

The present paper describes the capabilities of modern numerical methods for simulation of multiphase fluid flows in flow passages of centrifugal pumps, taking the package CFX-5.7 as an example. The paper suggests selection of different flow models, in accordance with the composition of a multiphase fluid and mode of its flow.

## Basic Notions of Multiphase Flows

First of all, we will distinguish between multicomponent and multiphase flows. Fluid medium is considered multicomponent if its substances are mixed at a microscopic, i.e., molecular level. Flow of a multicomponent mixture has a common field of velocity, temperature, etc., for all the components of the mixture. In multiphase flows, fluid media are mixed at a macroscopic level. Different fractions may have different physical properties and be simulated by different equations of motion and boundary conditions.

If the latter is the case, then such multiphase flow is considered to be inhomogeneous. In this case, different fractions have different fields of velocities and other variables, but the pressure field is common for all fractions of the mixture. In some cases, multiphase flows may be considered homogeneous. It means that all the fractions have common fields of velocity, temperature, pressure, turbulence, etc. Homogeneous flows are simulated by a common set of equations of momentum, energy, etc., for all the phases. Flows with explicit interface between phases (free surface flows) usually belong to this type.

As morphology of the substance of each phase, one of the following states can be specified. Continuous medium is a liquid or gas that forms continuous simply connected domain (water, air, etc.). Dispersed medium is a substance that is presented in form of discrete domains. The dispersed medium may consist of gas or liquid (water droplets in air, air bubbles in water) or solid particles (grains of sand in water, specks of dust in air). More complex case, poly-dispersed medium, includes dispersed phases with different dimensions of particles (e.g., milk includes fatty balls of different average diameters, from 0.1 to 10 μm).

## Simulation of Liquid Flows in the Flow Passage of Centrifugal Pumps in Case of Small Contents of Gas and/or Solid Particles

When the gas contents is low, gas fraction is present in the fluid flow in form of bubbles and does not exert significant influence on the flow pattern. As the gas contents grows, the head, capacity and consumed power of a pump decrease due to change of density of the gas-liquid mixture. For simulation of such flow, one of the following approaches may be used.

### 1. Lagrangian Approach

In this approach, called also Particle Transport Model, motion of separate dispersed particles is modeled, under influence exerted by the forces from the flow of the carrying phase. It is assumed for simplicity that dispersed particles are of ball shape. The forces that act on a particle are caused by difference of velocities of the particle and the flow of the carrying phase, as well as by displacement of the carrying medium by this particle. The equation of motion of a particle looks as follows:

$$m_p \frac{dv_p}{dt} = 3\pi\mu d C_{cor}(v_f - v_p) + \frac{\pi d^3 r_f}{6}\frac{dv_f}{dt} + \frac{\pi d^3 r_f}{12}\left(\frac{dv_f}{dt} - \frac{dv_p}{dt}\right) +$$

$$+ F_e - \frac{\pi d^3}{6}(r_p - r_f)\vec{w}\times(\vec{w}\times\vec{r}) - \frac{\pi d^3 r_p}{3}(\vec{w}\times v_p). \quad (1)$$

Here $m_p$ is the particle mass, $d$ is the particle diameter, $v$ is the velocity, $\mu$ is the dynamic viscosity of the fluid of main phase, $C_{cor}$ is its coefficient of viscous drag; $\vec{w}$ is the rotational speed, $r$ is the radius-vector (when considering flow in a relative frame of reference). Index $p$ refers to a particle; index $f$ refers to the fluid of the main phase.

The Eq. (1) is a first-order differential equation where the only unknown value is the particle velocity $v_p$, and the argument is the time $t$. Physical sense of separate terms of this equation is described, in particular, in [7].

At the preprocessor stage, a user specifies diameter and flow rate of the injected particles. Motion of each particle is modeled by own equation (1). Thus, the more particles are present in the computational domain, the more resource consuming will be the solution process.

This model permits also to simulate reverse influence exerted on flow of the carrying phase by motion of particles. In this case, after each injection of particles, the additional source term is computed for each particle. This term expresses influence of particles on the flow of the carrying phase and is added to the Navier – Stokes (Reynolds) equations for the flow of the carrying phase [8, 9].

*2. Eulerian Approach + Particle Model*

Each phase is assumed to be a continuous medium, and motion of the substance of each phase is simulated by own set of equations of Navier – Stokes (Reynolds), continuity and energy. In each set of equations, the terms that describe heat and mass transfer between phases are added. Values of these terms depend on the interphase contact area.

This approach can be used also in the case when one of the phases is present in form of dispersed particles (in particular, air bubbles in water) [8, 9]. In this case, the particle model should be specified for this phase, but the approach remains to be Eulerian. At the preprocessing stage, a user specifies diameter and flow rate of the injected particles. The Particle model is used only for evaluation of the contact area between phases, with the assumption that the particles are of spherical shape.

*3. Algebraic Slip Model*

At the preprocessing stage, a user specifies information about the multiphase mixture as single multicomponent substance with dispersed particles (e.g., water with grains of sand and air bubbles) [8, 9]. This multicomponent substance is assumed to be continuous medium, and its motion is simulated by a common set of Navier – Stokes (Reynolds), continuity and energy equations. It is assumed in this model that exchange of momentum between phases occurs only due to viscous friction. The shape of dispersed particles is assumed spherical. The equations of this model are described, e.g., in [7] (sub-section "Multiphase Mixture Model").

This model permits to take into account that different phases move with different velocities, using the notion of slip velocities. This allows for simulation, e.g., deceleration of sand grains flying into a vessel filled with still fluid. In other words, this model is appropriate when the trajectories of dispersed particles almost do not deflect from the streamlines of the carrier phase.

*4. Advantages and Disadvantages, Conclusion*

An advantage of the Lagrangian approach is obtaining of detailed information about behavior of separate particles of the dispersed phase in the flow of the carrier phase. However, this approach is inefficient when the number of particles is large. The Lagrangian approach is more efficient when the particles of different dimensions are present in the flow, whereas in the Eulerian approach the particles of different dimensions should be modeled as separate phases. At the same time, the Eulerian approach is more convenient for analysis of integral parameters of the flow.

Regarding the flow of multiphase fluid in the flow passage of a pump, we consider to be expedient to use the Lagrangian approach with the purpose to analyze behavior of separate particles. When the performance parameters of the pump at different composition of fluid is of interest, the Eulerian approach with the particle model is a good choice.

The Algebraic Slip Model has the same advantages as the Eulerian approach with the particle model but is less resource consuming. However, if the flow pattern does not correspond to the assumptions of this model, it will produce inaccurate results.

## Simulation of Liquid Flows in the Flow Passage of Centrifugal Pumps in Case of High Gas Contents

When the gas contents is high (for a free-vortex pump considered in [2], above 6% of fluid volume), gas bubbles in the fluid flow merge together and form the gas cavities. These cavities change substantially the flow pattern and exert influence on the performance parameters of the pump. In this case, both phases should be treated as continuos media, and for simulation of such flows different models of the Eulerian approach can be used.

*1. Free Surface Flows*

This approach allows for simulation of flow of two (or more) liquids or liquid with gas that don't intermix and, being under action of body forces, form a clear interface, i.e., free surface.

If the interface is clearly expressed, the flow may be simulated using a homogeneous model. In this case, the flow of both phases is modeled by a common set of continuity, momentum and energy equations. Besides, for approximation of the free surface, the model equations are supplemented with the equation of transfer of the fill function $F$ that expresses "concentration of liquid in gas" (when considering gas-liquid flow).

$$\frac{\partial F}{\partial t} + \frac{\partial}{\partial x_j}\left(F u_j\right) = 0. \qquad (2)$$

$F$ equals 1 in the volume occupied by liquid and 0 in the volume occupied by gas. Only for cells crossed by the free surface, $0 < F < 1$. Initial position of the free surface should be specified. Additional information regarding this model may be found, e.g., in [7].

If the interface is not clearly expressed, as the substance of one phase is involved in the substance of other phase, the inhomogeneous model is appropriate. For example, if in case of water-air flow significant splashing occurs, air bubbles can be entrained in water and then behave as a dispersed phase. Disadvantage of the inhomogeneous model is more resource consuming solution process, as motion of each phase is simulated in this case by the own set of differential equations.

*2. Mixture Model*

This approach allows to simulate flow of two (or more) liquids or gases that can interpenetrate and don't obligatory form the interface. This model is appropriate when none of the phases has dominant influence upon the flow pattern. An example is flow of water-oil mixture, when water bubbles in oil as well as oil bubbles in water may be present.

In both these models, the model equations for each phase are supplemented with terms that express heat and mass transfer between phases. The interphase contact area determines the values of these terms. For evaluation of this area (per volume unit), a user should specify the interphase length scale.

*3. Conclusions*

At the stage of preprocessing, a user should understand, at least approximately, what flow pattern he is going to simulate. According to the available experimental information on gas-liquid flows in the flow passage of a pump, high gas contents leads to formation of gas cavities with a clearly expressed free surface [2]. Besides, dispersion of air bubbles takes place. Its intensity depends on design of the pump and rotation speed of the rotor. Evidently, the best model for analysis of such a flow is the inhomogeneous free surface model or the mixture model. In some cases good results may be obtained also using the homogeneous model.

**Simulation of Other Physical Phenomena**

*1. Compressibility*

When considering flow of a gas-liquid mixture, the liquid is considered incompressible medium, whereas the gas may be compressed or expanded under action of pressure field in the computational domain [10]. In the flow passage of a pump, as the pressure increases from the inlet to the outlet, the gas is compressed. Besides, under the fixed pump head, the higher is pressure at the pump inlet, the stronger is gas compression. Compressibility of gas is simulated by switching on the energy equation [7].

*2. Buoyancy*

The buoyancy phenomenon is manifested as appearance of flow in the field of body forces due to non-uniform distribution of density, e.g., due to non-uniform heating. In multiphase media, a reason for emerging of buoyancy is also difference in density between the substances of different phases, as it is almost always essential.

In the flow passage of a pump, the body forces are centrifugal force and Coriolis force (when considering flow in a rotation frame of reference). Principle of operation of separating devices is based just on the buoyancy phenomenon. In these devices, the gas is separated from the liquid and accumulated in the low-pressure regions. Regarding the flow of the gas-liquid mixture in the flow passage of a pump, the gravity force may also be essential. In particular, according to [2], the limited gas contents in the flow passage of a pump depends on whether the pump is installed vertically or horizontally.

*3. Surface Tension*

The larger is surface tension, the more the gas-liquid mixture is inclined to form large stable gas bubbles and cavities, and the more difficult is to break these bubbles and cavities. The coefficient of surface tension is an important physical property of the medium. One of the ways to prevent collapse of the performance parameters of pumps treating gas-liquid mixtures is introducing of surfactant species that reduce this coefficient [11]. Modern software packages allow taking also this tension into account. A user specifies the coefficient of surface tension at the stage of preprocessing.

*4. Modeling of Turbulence*

In homogeneous flows, one and the same turbulence model is used for all the components. In inhomogeneous flows, different turbulence models may describe the motion of substances of different phases. However, in case of a dispersed phase (e.g., bubbles) (in the Eulerian approach), the flow of this phase is considered laminar or described by an eddy viscosity model, as application of more complex turbulence models is considered inexpedient in this case. In case of continuous phases, when using the free surface model or the mixture model, for each of the phases any complex turbulence model may be used [7]. As a rule, it is expedient to use one and the same turbulence model for each of the phases.

*5. Heat and Mass Transfer between Phases: Heating, Evaporation, Dissolving, Cavitation*

All these processes can also be simulated using modern software packages [8, 9]. This is achieved by adding of additional terms to the momentum equations. The selected model determines the appearance of these terms.

Regarding the flow in the flow passage of a centrifugal pump, in some cases, the cavitation phenomenon is of large importance. At the preprocessing stage, a user specifies that in the computational domain, e.g., water (liquid) and water steam (gas) are present. At the initial moment, the concentration of water steam is assumed to be zero. It becomes above zero when the pressure in some locations of the flow passage drops below the saturation pressure.

In particular, in the package CFX-5.7, the cavitation model of Rayleigh – Plesset is implemented. An example of application of this model is presented in [12].

*6. Clogging*

When computing flows with deposition of solid particles at the walls, it is possible to specify the maximum volume fraction that can be occupied by the solid substance (Maximum Packing) [9]. If the particles are of spherical shape, this coefficient equals 0.74.

*7. Transient Effects*

As it is described in the paper [6], for sparing of computational resources, numerical simulation of flows inside the pumps is performed usually in a steady state formulation. From the other side, considering multiphase flows, investigation of some transient processes may be of interest. In particular, an interesting phenomenon is the process of rupture of continuous fluid flow due to high gas contents (for a free-vortex pump considered in [2], above 40% of volume of mixture). In this case, at the interface between a rotating impeller and stationary components, the option of Transient Rotor-Stator should be specified [6].

## Conclusions

The article describes the flow models and suggests recommendations on their selection for prediction of flow pattern in the flow passage and performance parameters of centrifugal pumps treating multiphase mixture.

*Acknowledgements*

The present research was conducted under leadership of Asst. Prof. A.A. Yevtushenko and under support of the collective of the department of applied fluid mechanics.